\documentclass[11pt,twoside]{article}  
\usepackage{apn3conf}

\begin{document}   

%
%
%
%

\title{Evidence for bipolar jets in late stages of AGB winds}

%
%
%

\author{D. Vinkovi\'{c}\altaffilmark{1}, M. Elitzur}
\affil{Department of Physics \& Astronomy, University of Kentucky,
Lexington, KY 40506-0055}
\author{K.-H. Hofmann, G. Weigelt}
\affil{Max-Planck-Institut f\"{u}r Radioastronomie, Auf dem H\"{u}gel 69, 53121
Bonn, Germany}

\altaffiltext{1}{School of Natural Sciences, Institute for Advanced Study, Princeton, NJ 08540}
%
%

\contact{Dejan Vinkovic}
\email{dejan@pa.uky.edu}

%
%
%
%
%

\paindex{Vinkovi\'{c}, D.}
\aindex{Hofmann, K.-H.}
\aindex{Elitzur, M.}
\aindex{Weigelt, G.}     

%
%

\authormark{Vinkovi\'{c}, Hofmann, Elitzur \& Weigelt}

%
%

\keywords{dust, radiative transfer, IRC+10011, AGB stars}


\begin{abstract}          
Bipolar expansion at various stages of evolution has been recently observed in
a number of AGB stars. The expansion is driven by bipolar jets that emerge late
in the evolution of AGB winds. The wind traps the jets, resulting in an
expanding, elongated cocoon. Eventually the jets break-out from the confining
spherical wind, as recently observed in W43A. This source displays the most
advanced evolutionary stage of jets in AGB winds. The earliest example is
IRC+10011, where the asymmetry is revealed in high-resolution near-IR imaging.
In this source the jets turned on only $\sim$ 200 years ago, while the
spherical wind is $\sim$ 4000 years old.
\end{abstract}

%
%

The premise that asymmetric planetary nebulae (PNe) evolve from spherically
symmetric Asymptotic Giant Branch (AGB) winds raises one of the most intriguing
questions of stellar astrophysics: when and how does the stellar wind break its
spherical symmetry? Since many proto-planetary nebulae (PPNe) show distinct
large-scale asymmetries (e.g. Su, Hrivnak \& Kwok 2001, Sahai 2002), the
symmetry breaking process has to operate already during the late stages of AGB
evolution. This implies that AGB winds are not always spherical. Indeed, in
recent years, high-resolution imaging has yielded conclusive evidence of
asymmetry for several AGB objects (V Hya: Plez \& Lambert\ 1994, Sahai et al.\
2003; X Her: Kahane \& Jura 1996; IRC+10011: Hofmann et al.\ 2001; IRC+10216:
Weigelt et al.\ 1998 \& 2002, Haniff \& Buscher 1998, Skinner et al.\ 1998,
Osterbart et al.\ 2000; RV Boo: Bergman et al.\ 2000, Biller et al.\ 2003;
CIT6: Lindqvist et al.\ 2000, Monnier et al.\ 2000, Schmidt et al.\ 2002).
There have been suggestions that asymmetries could be even prevalent in AGB
winds (Plez \& Lambert\ 1994, Kahane at al.\ 1997).

\section{Jet-driven asymmetry evolution in AGB stars}

Concurrently, an increasing number of jet and jet-like features has been
identified in various PNe and PPNe, prompting a suggestion that jets are also
responsible for symmetry breaking in AGB winds (Sahai \& Trauger 1998). The
strongest evidence for jets comes from maser observations, including proper
motion measurements, of both the fast collimated jets and the slow spherical
wind (e.g., IRAS 16342-3814: Sahai et al.\ 1999; K3-35: Miranda et al.\ 2001;
Hen 3-1475: Riera et al.\ 2003; IRAS 22036+5306: Sahai et al.\ 2003). Probably
the youngest display of such a configuration of masers is the AGB star W43A
(Imai et al.\ 2002). Considering that these jets are detected when they already
break out from the confinement of a slow high density AGB wind, there has to
exist an earlier instance of jet-evolution when only a small expanding cocoon
is detectable (Scheuer 1974).

If large scale bipolar jet-cavities were to be carved out from the AGB wind of
PPNe, as demonstrated in IRAS 16342-3814, then the cocoon expansion has to
operate already during the AGB phase. Thus the aforementioned asymmetric AGB
objects should provide glimpses of the cocoon expansion. The direct kinematic
evidence of this process exists only for V Hya and W43A, where expansion
velocities are measured. In the other objects, asymmetries revealed in imaging
observations fit into this scenario.

The youngest example of a jet-cocoon in AGB stars, as we argue in
section \ref{CIT3cocoon}, exists in IRC+10011. A more advanced
stage of cocoon evolution is present in the prototype C-rich star
IRC+10216. Its cocoon is of a similar size, but the asymmetry is
more pronounced and evident even in the K-band, unlike IRC+10011
where only the J-band image shows clear asymmetry. The C-rich star
V Hya provides an example that is further along in evolution.
Recent CO observations by Sahai et al.\ (2003) show a morphology
of a bipolar outflow with velocities of 100--150 \hbox{km
s$^{-1}$} breaking from the confinement of the high-density region
of the slowly expanding AGB wind. A similar structure has been
found in the O-rich star X Her, with a spherical wind of 2.5
\hbox{km s$^{-1}$} and two symmetrically displaced 10 \hbox{km
s$^{-1}$} components, likely to be the red and blue shifted cones
of a weakly collimated bipolar flow. An even more evolved system
is the C-rich star CIT6, where a bipolar asymmetry dominates the
image both in molecular line mapping and in HST-NICMOS imaging.

These examples show that a bipolar asymmetry, most probably created by
collimated outflows, appears during the final stages of AGB mass outflow. This
represents the first instance of symmetry breaking in the evolution from AGB to
planetary nebula. It is still not clear, however, what physical process drives
these jets. Diversity in their properties could lead toward diversity in
geometrical and physical properties of the bipolar asymmetry. When the AGB
phase ends, a mixture of various processes emerges, such as multiple jets and
fast winds. Their interaction with the AGB circumstellar asymmetries leads to
the myriad of complex structures found in PPN and PN sources.

\section{IRC+10011: the youngest example of jets in AGB stars?}

\begin{figure}
 \epsscale{.32}
 \plotone{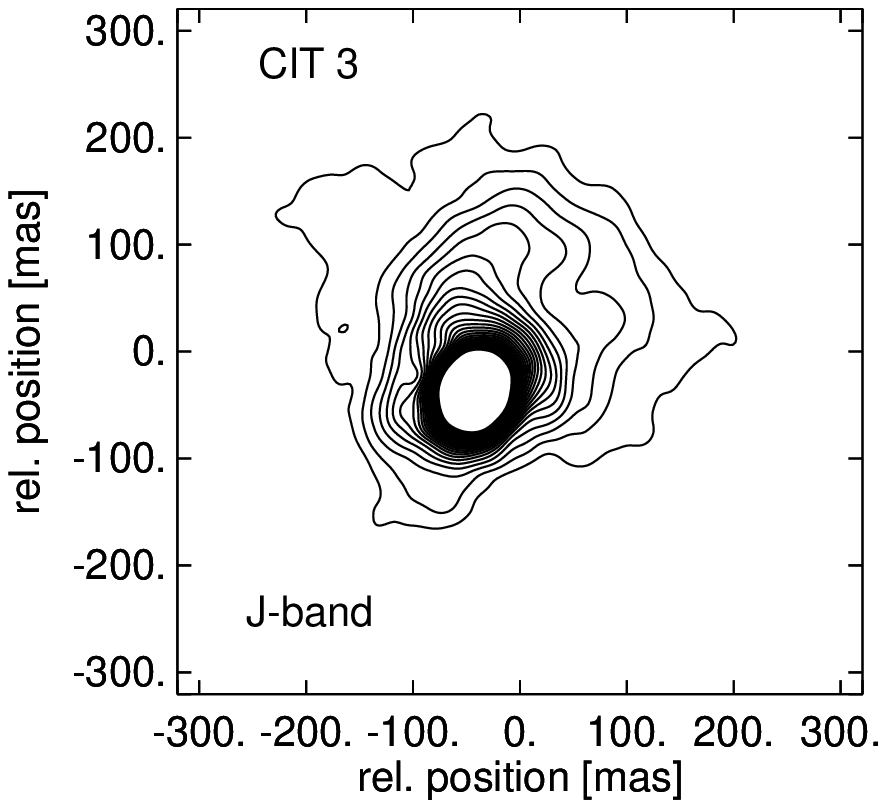}
 \plotone{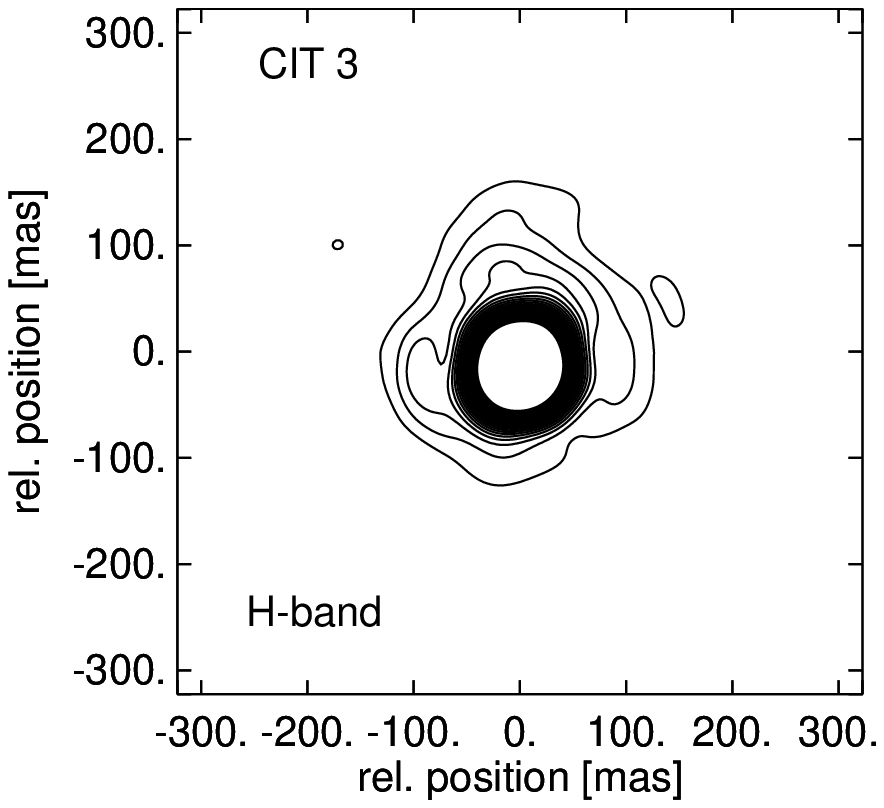}
 \plotone{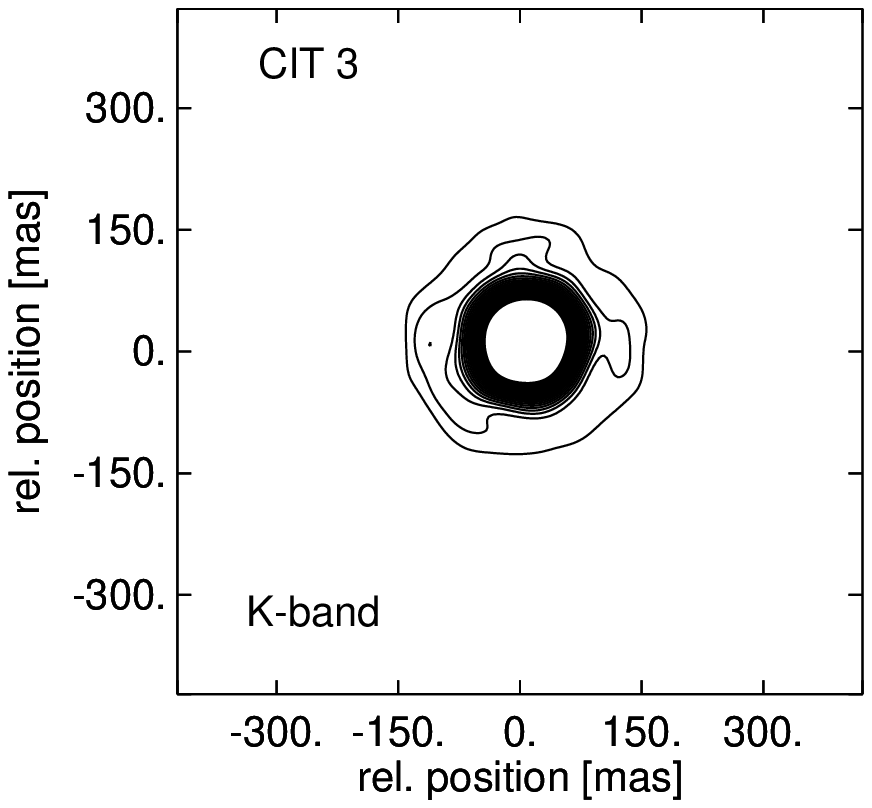}
 \epsscale{.29}
 \hspace{0.5cm}
 \plotone{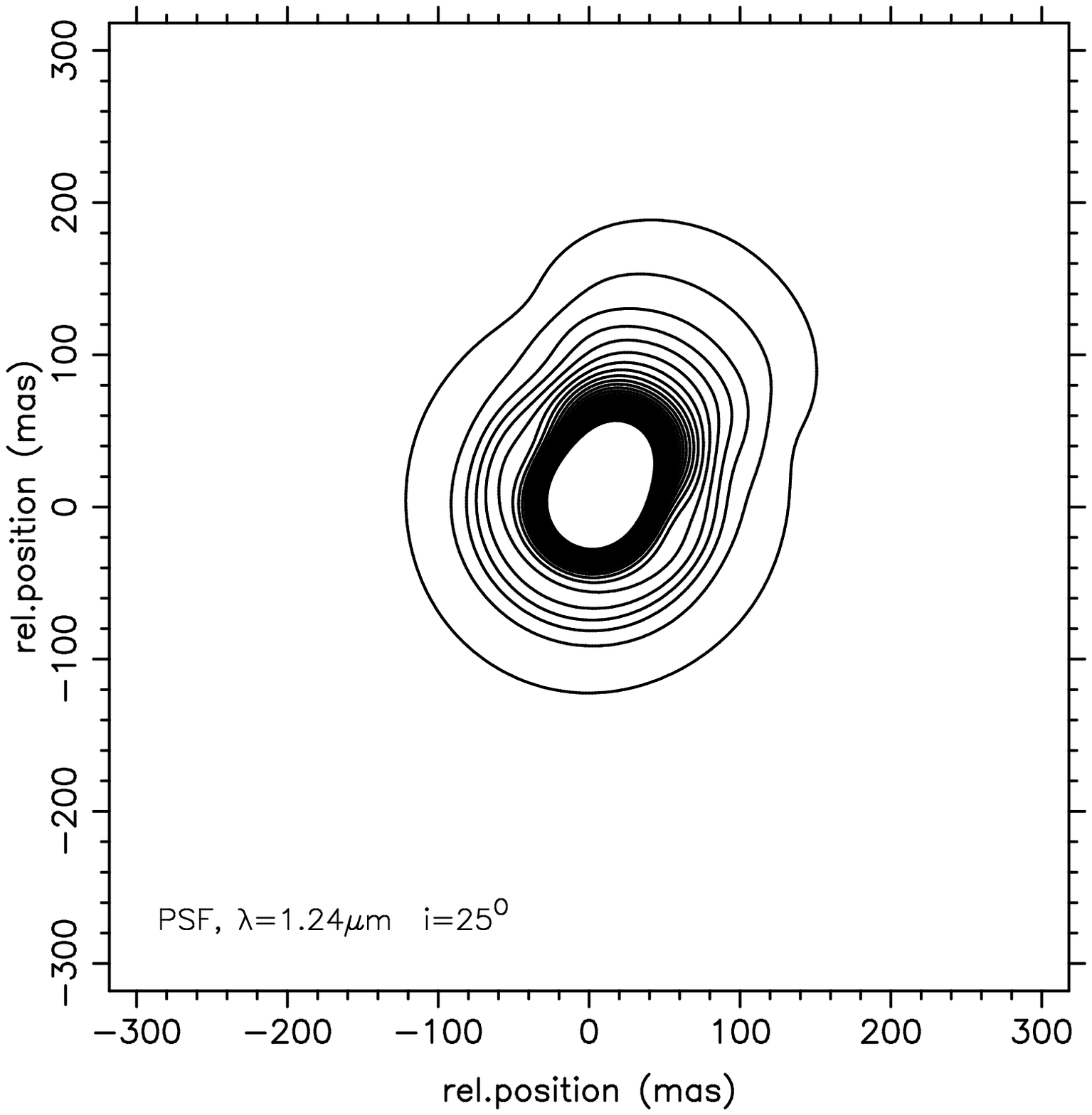}
 \hspace{0.4cm}
 \plotone{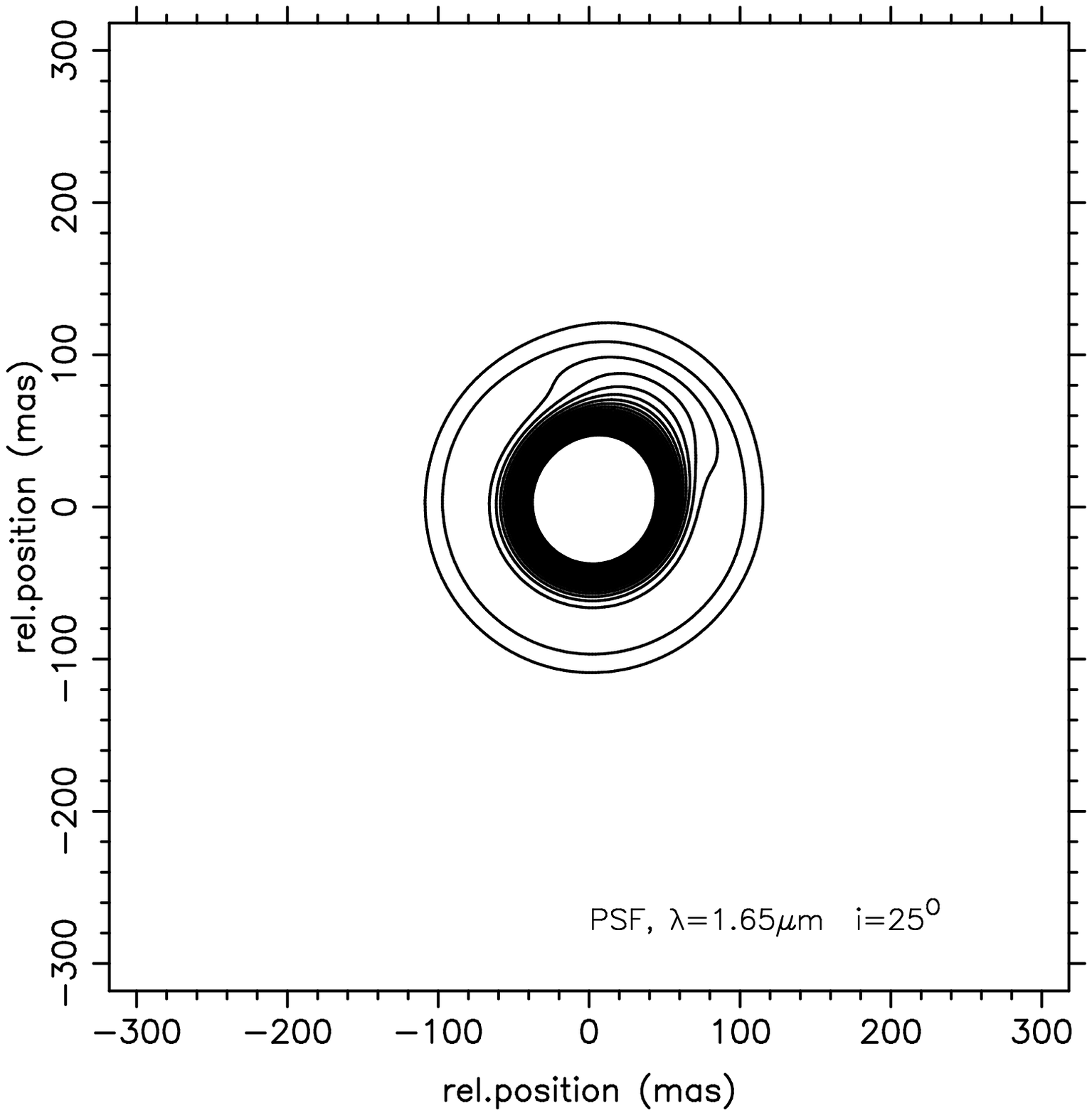}
 \hspace{0.4cm}
 \plotone{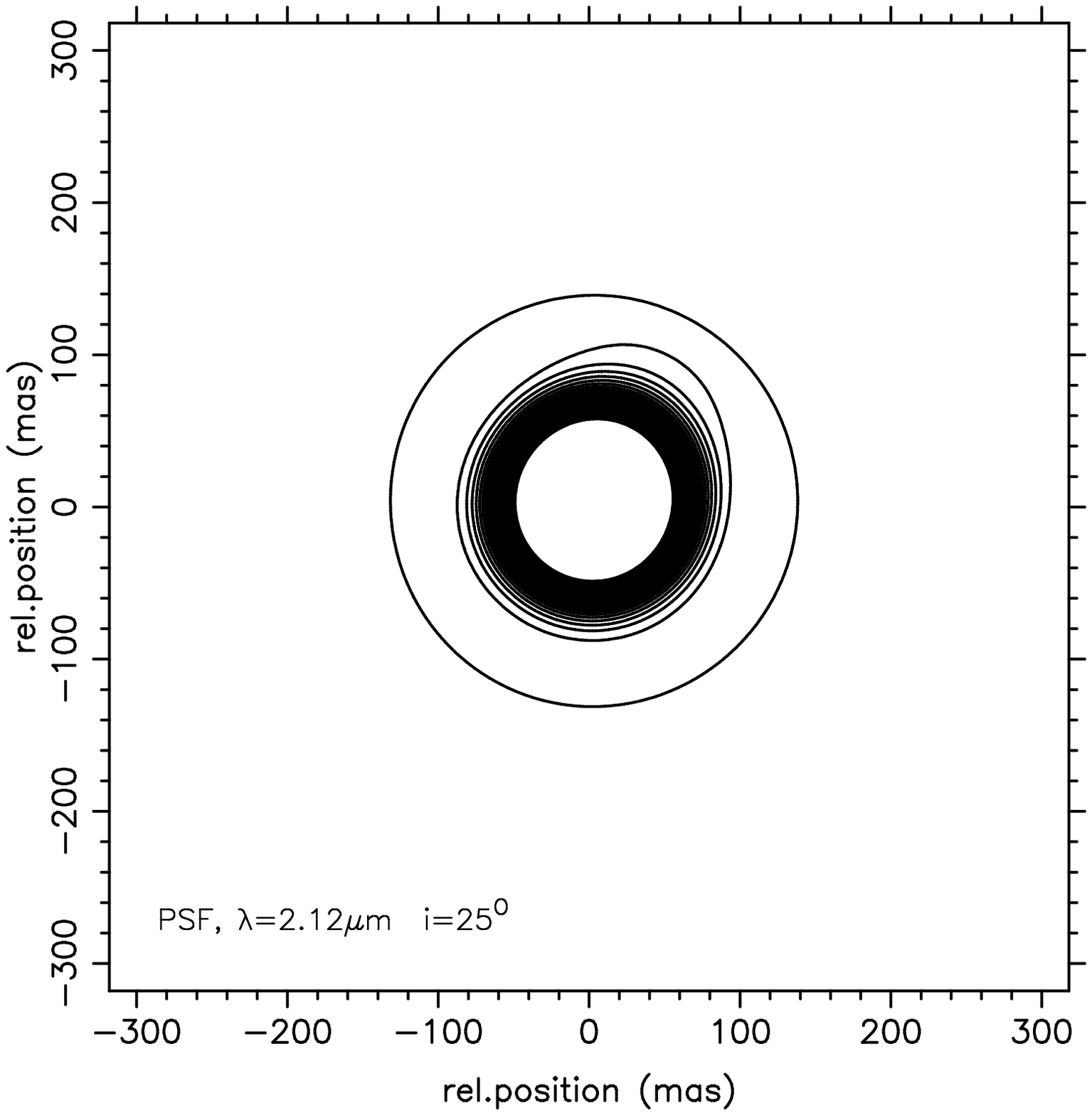}
\caption{Upper row: Near infrared images of IRC+10011 in the J-
(1.24$\mu$m), H- (1.65$\mu$m) and K'-band (2.12$\mu$m) with
respective resolutions of 48 mas, 56 mas and 73 mas (H01). Lower
row: Theoretical images based on 2D radiative transfer convolved
with the instrumental PSF of H01. Contours are plotted from 1.5\%
to 29.5\% of the peak brightness in steps of 1\%. The transition
from scattered light dominance in the J-band to thermal dust
emission in the K'-band creates a sudden disappearance of the
image asymmetry (Figure \ref{CIT3sketch}). The bright fan-shaped
structure is scattered light escaping through the polar cone
carved out by the jet.} \label{CIT3images}
\end{figure}

The oxygen-rich star IRC+10011 (= IRAS 01037+1219 = CIT3 = WXPsc)
is one of the most extreme infrared AGB objects. It is surrounded
by an optically thick dusty shell formed by a large mass loss rate
of $\sim$\ 10$^{-5}$M$_\odot$yr$^{-1}$. The shell expansion
velocity of $\sim$\ 20 \hbox{km s$^{-1}$} has been measured in OH
maser and CO lines. This source served as the prototype for the
first detailed models of AGB winds by Goldreich \& Scoville\
(1976) and of the OH maser emission from OH/IR stars by Elitzur,
Goldreich, \& Scoville\ (1976). The sphericity of its dusty wind
had been considered secure. The recent discovery by Hofmann et
al.\ (2001; H01 hereafter) of distinct asymmetries in its envelope
was, therefore, a surprise. They obtained the first near infrared
bispectrum speckle-interferometry observations of this source with
a resolution of tens of AU (Figure \ref{CIT3images}, upper row).
While the H- and K'-band images appear almost spherically
symmetric, the J-band shows a clear asymmetry.

\subsection{2D radiative transfer modeling}

\begin{figure}
 \epsscale{.39}
 \plotone{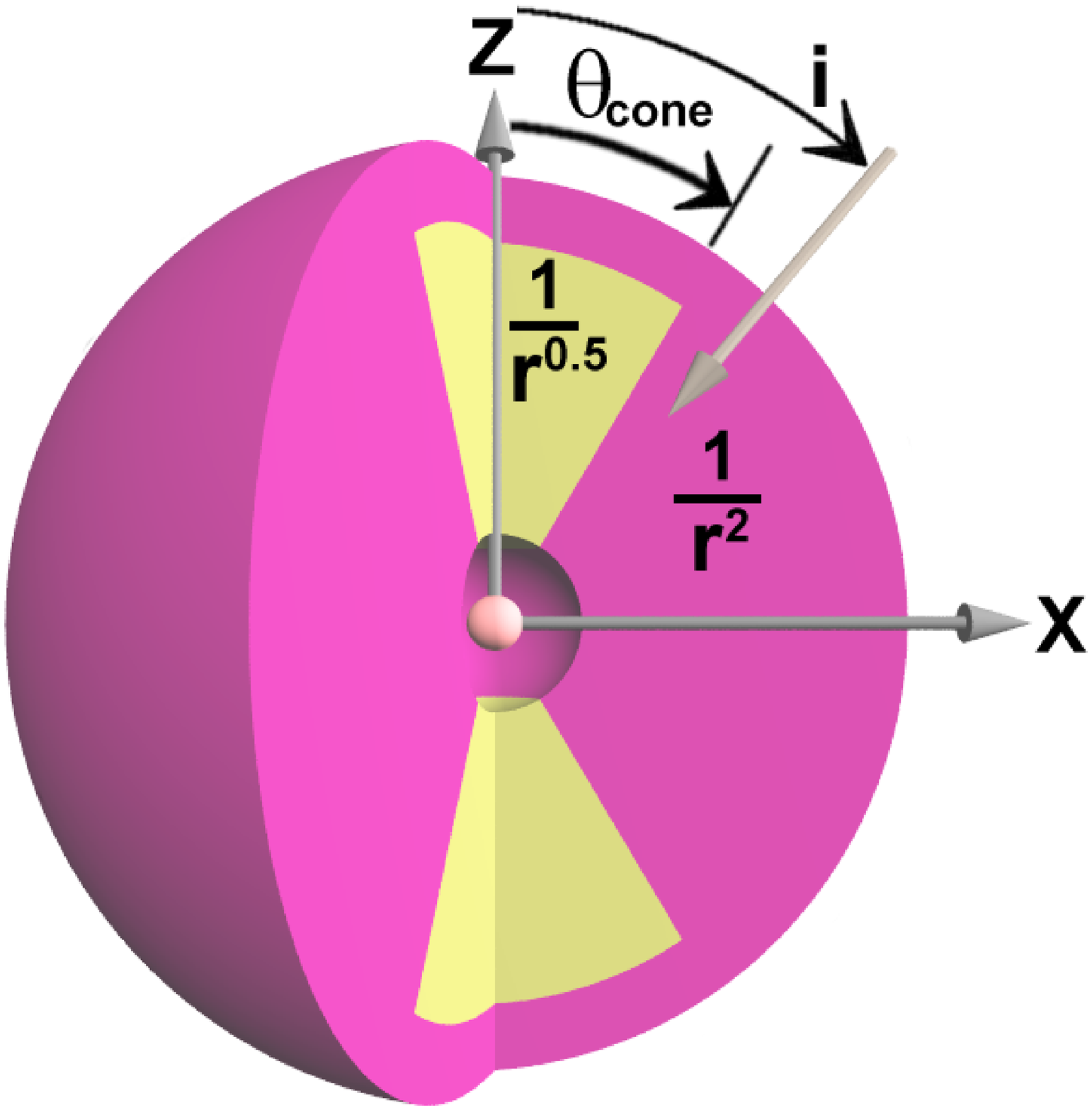}
 \epsscale{.59}
 \plotone{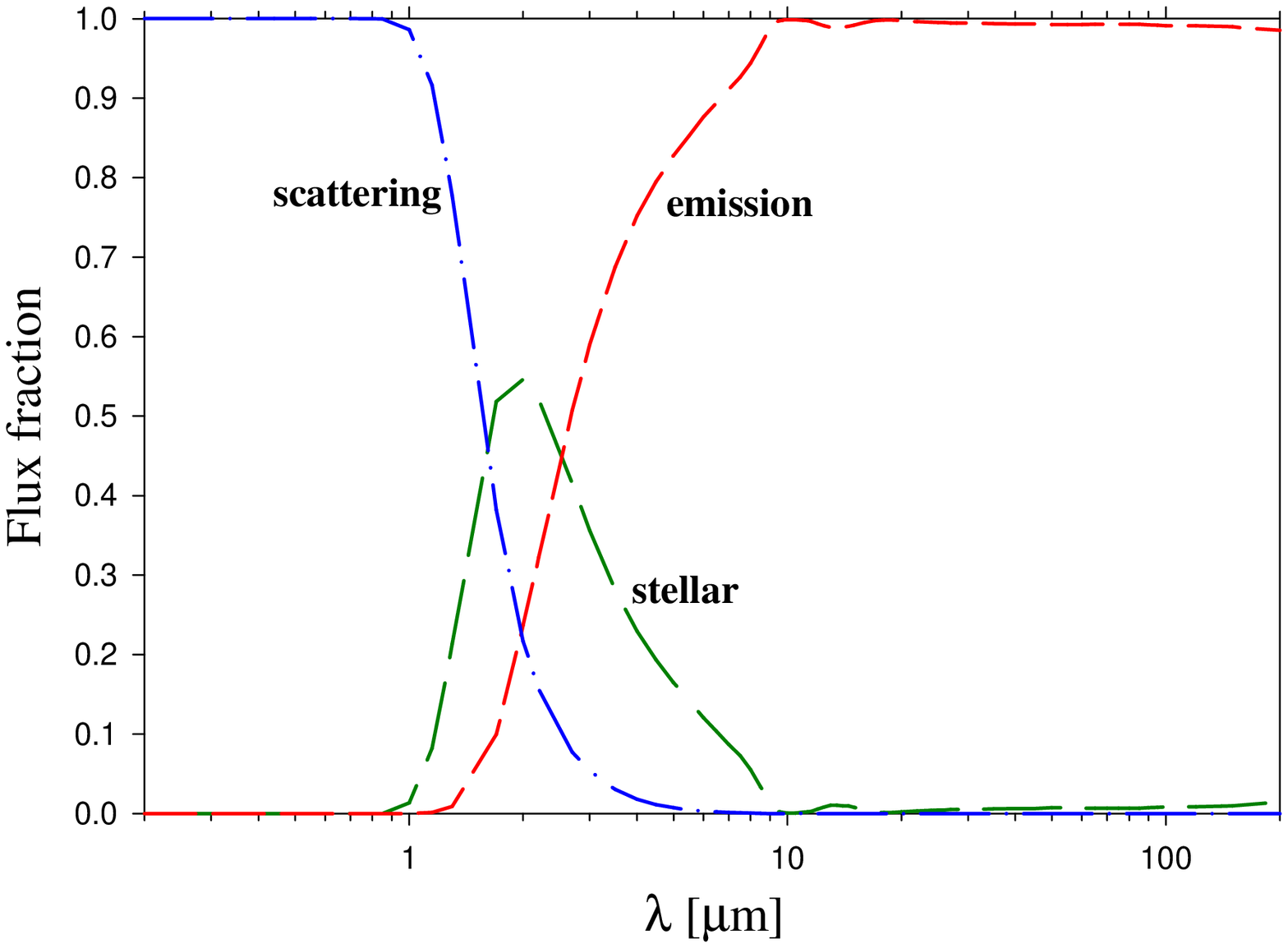}
 \caption{Left: Sketch of the 2D model for the circumstellar dusty shell
around IRC+10011. Two polar cones with half-opening angle
$\theta_{cone}$ and a $1/r^{0.5}$ density profile are imbedded in
a spherical wind with the standard $1/r^2$ density profile. The
system is viewed from angle $i$ to the axis. Right: Wavelength
variation of the relative contribution of each component to total
flux. Note the fast change from scattering to emission dominance
around 2$\mu$m. This transition is responsible for the observed
wavelength variation of the image asymmetry in the near-IR.}
\label{CIT3sketch}
\end{figure}

These images impose strong constrains on the circumstellar dust
density distribution in the inner regions. It can be easily shown
that scattering by a $1/r^p$ dust density distribution produces a
$1/r^{p+1}$ brightness profile. Since the J-band image at 1.24
$\mu$m is dominated by dust scattering, the following geometry was
deduced directly from the image: along the polar axis, the dust
density declines as $1/r^{0.5}$, while the rest of the dusty wind
has the typical $1/r^2$ density profile (Figure \ref{CIT3sketch},
left). Without good observational constraints on the large scale
structure, we adopted the approach of the {\it minimum number of
free parameters} in which the bipolar cones radially extend to
$R_{\rm cone}$ with the half-opening angle $\theta_{\rm cone}$.
Thus our model should be considered only a first-order
approximation to the actual structure of IRC+10011.

The radiative transfer calculation was performed with LELUYA, a 2D
code that can handle arbitrary axially symmetric dust
configurations without approximations (www.leluya.org). General
scaling properties of the radiatively heated dust (Ivezi\'{c} \&
Elitzur 1997) have been applied such that the dust condensation
temperature of 900 K is the only dimensional quantity specified.
All other properties can be expressed in dimensionless terms.
Luminosity is irrelevant, only the stellar spectral shape is
needed, which we take as black-body at 2,250 K. Similarly, only
the spectral shapes of the dust absorption and scattering
coefficients are needed. We use the silicate grains of Ossenkopf,
Henning \& Mathis (1992) with the standard size distribution
described by Mathis, Rumpl \& Nordsieck (1977). However, we found
that fits of the visibility curves require the upper limit on the
grain sizes reduced to 0.20 $\mu$m from the standard 0.25 $\mu$m.

The best fit to the overall spectral energy distribution and
imaging visibility curves yields visual optical depth of 20 along
the conical symmetry axis and 40 in the equatorial plane. The
angular radius of the dust condensation cavity is $\sim$ 35 mas
(23 $\pm$ 5 AU).  The fit also yields a bolometric flux of
$10^{-9}\,\rm W/m^2$, corresponding to 10.82~mas for the stellar
angular size. The system is observed at an inclination
$i=25^{\circ}\pm 3^{\circ}$ from the axis so that the wind
obscures the receding part of the bipolar structure, creating the
observed asymmetry of the scattering image. When imaged at long
wavelengths, such as the K'-band, the dust thermal emission starts
to dominate (Figure \ref{CIT3sketch}, right). The asymmetry is
then less apparent since the central heating by the star tends to
produce spherical isotherms.

Assuming standard gas-to-dust ratio, the overall mass of the IRC+10011
circumstellar shell is 0.13M$_\odot$. The measured wind velocity of 20
\hbox{km\ s$^{-1}$} gives a lifetime of $\sim$ 3,800 years for the current
outflow phase. This corresponds to a mass loss rate of
3$\times$10$^{-5}$M$_\odot$yr$^{-1}$, which is a signature of the very end of
AGB evolution. Our modeling suggests the existence of a toroidal density
enhancement at distances of more than $\sim$ 100 AU. Imaging observation by
Marengo et al.\ (1999) provide a possible indication of such structure.

\subsection{Bipolar cones as expanding jet cocoons}
\label{CIT3cocoon}

The derived gas density at the base of each cone is 1.3$\times$10$^6$cm$^{-3}$,
while at the base of the wind region it is 1.7$\times$10$^8$cm$^{-3}$. This
large disparity implies that the bipolar cones are sustained by high-velocity
ram pressure, most probably due to high-velocity low-density jets. We find that
the opening angle of the cones is 2$\theta_{\rm cone} = 30^{\circ}$ and they
extend to about 500 to 1000 AU. The cone emission comes from the swept-up
ambient wind material. All of these properties are consistent with the jet
evolution picture first described by Scheuer (1974): jets are clearing the
polar regions but are trapped by the ambient material pushed ahead by their ram
pressure, resulting in an expanding cocoon. The current density distribution in
the cocoon is just a snapshot of an inherently dynamic structure. We find the
jet lifetime is $\sim$ 200 years. In comparison with the other asymmetric AGB
objects, IRC+10011 currently provides the youngest known example of asymmetry
in AGB outflows.

%
%
%
%


\end{document}